\documentstyle[12pt]{article}

\newfont{\Mb}{msbm10}

\begin{document}
\setcounter{equation}{0}
\setcounter{figure}{0}
\setcounter{table}{0}

\hspace\parindent
\thispagestyle{empty}

\bigskip
\bigskip
\bigskip
\begin{center}
{\LARGE \bf A Semi-Algorithmic Search}
\end{center}
\begin{center}
{\LARGE \bf for Lie Symmetries}
\end{center}

\bigskip

\begin{center}
{\large
L.G.S. Duarte and L.A.C.P. da Mota
\footnote{E-mails:
lduarte@dft.if.uerj.br and damota@dft.if.uerj.br}
}
\end{center}

\bigskip
\bigskip
\centerline{\it Universidade do Estado do Rio de Janeiro,}
\centerline{\it Instituto de F\'{\i}sica, Depto. de F\'{\i}sica Te\'orica,}
\centerline{\it 20559-900 Rio de Janeiro -- RJ, Brazil}

\bigskip
\bigskip

\bigskip
\bigskip

\abstract{In \cite{Nosjpa2001} we defined a function (we called $S$) associated
to a rational second order ordinary differential equation (rational 2ODE) that
is linked to the search of an integrating factor. In this work we investigate
the relation between these $S$-functions and the Lie symmetries of a rational
2ODE. Based on this relation we can construct a semi-algorithmic method to find
the Lie symmetries of a 2ODE even in the case where it presents no Lie point
symmetries.}

\bigskip
\bigskip
\bigskip
\bigskip
\bigskip
\bigskip



\newpage

\section{Introduction}
\label{intro}

Lie group analysis is perhaps the most powerful tool to study
differential equations (DEs) \cite{Lie,Olver,Olver2,Ibragimov,Stephani,Bluman,
Bluman2,Dresner,Cantwell}.
S. Lie \cite{Lie} demonstrated that the majority of the technics of
solving differential equations could be unified under the same theoretical
background: the invariance of the DEs being solved under a continuous group
of transformations (a Lie group). Since its appearance in the end of the
19th century, the Lie analysis of DEs has suffered a fantastic grown up
especially in the last decades. One of the main reasons is, probably, the
appearance of the computer in the second half of the 20th century, since
it turns long and complicated symbolic calculations into the simple pushing
of a button. Unfortunately, in real life, nothing is that simple. For the
computer have the job done, we need to `tell' it exactly what to do. In
other words, it is necessary to furnish an algorithm, i.e., a finite sequence
of determined steps. However, in Lie's method, there is no algorithmic
procedure to solve the determining equations for the {\em infinitesimals}
(i.e., the coefficients of the symmetry generators)\footnote{Many computer
packages implement some {\it heuristics} to find the infinitesimals and apply
the Lie method. See, for example, \cite{Noscpc11997,Noscpc21998}.}. The things
go worse when we are dealing with an ODE not possessing Lie point symmetries
(presenting only dynamical symmetries). In this case we can not separate the determining equation in the derivatives and, instead of determining equations,
we have only one determining equation: a `very ugly' partial differential
equation (PDE) that may leave us completely lost. In this last situation,
we can not count even with a sistematic way to search for the symmetries; and,
witout the symmetries, the Lie method can not be applied.

To overcome the difficulties in the treatment of ODEs (or systems of ODEs) that
do not possess Lie point symmetries several approaches have been developed:
P. J. Olver introduced the concept of {\em exponencial vector field} (see
\cite{Olver}, p. 185); B. Abraham-Shrauner, A. Guo, K.S. Govinder, P.G.L. Leach,
F.M. Mahomed, A.A. Adam and others worked with the concept of {\em hiden} and
{\em non local symetries} \cite{AbGu,AbGu2,AbGoLe,Abraham,GoLe,AdMa}.
C. Muriel and J.L. Romero have developed the concept of {\em $\lambda$-symmetry}
\cite{MuRo,MuRo2} and E. Pucci and G. Saccomandi created the concept of
{\em telescopical symmetry} \cite{PuSa}. Another great approach was brought by
M.C. Nucci by making use of the {\em Jacobi last multiplier} \cite{Nucci}.

Despite all these developments, there is still no fully algorithmic method
to solve the determining equations. This mean that, in some part of the whole
process, we can face a set of PDEs (the determining equations) for the
infinitesimals we don't know how to deal with.

In \cite{Nosjpa2001} we have developed an extension of the Prelle-Singer
method \cite{PrSi} and, in that paper, we have proposed to use an
unknown function (that we called $S$ \footnote{This idea was further
pursued by \cite{chandra1,chandra2,chandra3}.}) in order to make the 1-form\footnote{This
1-form is associated with the rational 2ODE $y''=\phi(x,y,y')$, where
$\phi$ is a rational function of $(x,y,y')$.} $\,\phi(x,y,y')\,dx-dy'\,$
proportional to an exact 1-form. In \cite{Nosamc2007}, we constructed a
semi-algorithm to determine the $S$-function for an 2ODE presenting an
elementary\footnote{For a formal definition of elementary function,
please see \cite{Davenport}.} first integral. In this paper we study the
relation between the $S$-functions and the Lie symmetries of a rational
2ODE. Based on this relation we propose a semi-algorithm to calculate
the Lie symmetries for rational 2ODEs. This procedure can succeed even
in the case where there are only dynamical symmetries (i.e., no Lie
point symmetries).

In this work the main idea is to present a connection between a Darboux
type method and the search for Lie symmetries of 2ODEs. Here, we present
one such algorithm that we expect is the firt of a series of developments in that area.

The paper is organized as follows:
In section \ref{lieands}, we present some basic definitions and stablish the
relation between the $S$-functions and the Lie symmetries of a 2ODE.
In section \ref{calcls}, we propose a method (a semi-algorithm) to calculate
a Lie symmetry {\em associated} with an $S$-function.
In section \ref{examples}, we present two examples to show the method in
action.
Finally, we present our conclusions and point out some directions to further
our work.

\section{Lie Symmetries and $S$-functions}
\label{lieands}

In this section, we will describe how the $S$-function is defined and the
relation between the Lie symmetries and the $S$-functions of a 2ODE.

\subsection{The definition of the $S$-functions}
\label{defs}

Let us consider the 2ODE given by:

\begin{equation}
\label{2ode}
y'' = \phi(x,y,y'),
\end{equation}
where $\phi$ is a function of $(x,y,y')$.

A function $I(x,y,y')$ defines a first integral (conserved quantity) of
(\ref{2ode}) if $I(x,y,y')$ is constant over all solution curves of
(\ref{2ode}). So, in other words, the 1-form defined by
$\omega \equiv dI = I_x\,dx+I_y\,dy+I_{y'}\,dy'$ is null if the 1-forms
($\alpha,\,\beta$) defined by:
\begin{eqnarray}
\alpha &=& \phi\,dx - dy' \nonumber \\
\label{alfabeta}
\beta & =& y'dx - dy
\end{eqnarray}
\noindent
are null. This implies that

\begin{equation}
\label{di}
\omega = dI = r\alpha\,+\,s\beta
\end{equation}
\noindent
where $r$ and $s$ are functions of $(x,y,y')$.
Thus
\begin{equation}
\label{di2}
 dI=I_xdx+I_ydy+I_{y'}dy = r(\phi\,dx - dy')\,+\,s(y'dx - dy)
\end{equation}

\noindent implying that $\,I_x = r\,\phi+s\,y',\,\,I_y = - s\,$
and $\,I_{y'} = - r$. Therefore, if we determine $r$ and $s$, we
can find $I$ via quadratures (see \cite{Nosjmp2009,Nosjpa2010}).

Let us make some definitions that make it easier to show some connection
between the coeficients $r$ and $s$ and the symmetries of the 2ODE (\ref{2ode}).

\bigskip
\bigskip
{\bf Definition 1:} {\it A function $R(x,y,y')$ satisfying
\begin{equation}
\label{Rfunc}
R\,(A\,dx + B\,dy + C\,dy') = d\,\gamma ,
\end{equation}
\noindent
where $A,\,B\,$ and $\,C\,$ are functions of $(x,y,y')$ and $d\,\gamma$ is
an exact 1-form, is called an integrating factor for the 1-form
$(A\,dx + B\,dy + C\,dy')$.}

\bigskip
\bigskip
{\bf Definition 2:} {\it Let $S(x,y,y')$ be a function defined by
\begin{equation}
\label{Sfunc}
S \equiv \frac{s}{r},
\end{equation}
\noindent
where $\,r\,$ and $\,s\,$ are functions satisfying (\ref{di}).
We will call it a $S$-function associated with the 2ODE (\ref{2ode}).}

\bigskip
\noindent
Rewriting (\ref{di2}), we have
\begin{equation}
\label{di3}
dI=I_xdx+I_ydy+I_{y'}dy = r\left((\phi+\,\frac{s}{r}\,y')dx -
\frac{s}{r}\,dy - dy'\right).
\end{equation}
\noindent
We can see that $\,r\,$ is an integrating factor for the 1-form
$\left((\phi+\,\frac{s}{r}\,y')dx - \frac{s}{r}\,dy - dy'\right)$.
So, writing $\,r=R\,$ and using the definition for the $S$-function we can
finally write
\begin{equation}
\label{di4}
dI=R\left((\phi+\,S\,y')dx - S\,dy - dy'\right).
\end{equation}

\subsection{The connection of the $S$-functions with the Lie symmetries of a
2ODE}
\label{connectsl}

We will begin this section by stating a theorem (our first goal):

\bigskip
\bigskip
{\bf Theorem 1:} {\it Let $\,y''=\phi(x,y,y')\,$ be a 2ODE presenting
a first integral $\,I(x,y,y')\,$. If $\,\overline{\eta}(x,y,y')\,$ is the
infinitesimal of a Lie symmetry generator in the evolutionary form, then the
function $\,S\,$ defined by
\begin{equation}
\label{sfunc}
S \equiv - \frac{D_x[\overline{\eta}]}{\overline{\eta}},
\end{equation}
\noindent
where $\,D_x\,$ is the operator defined by
\begin{equation}
\label{defd}
D_x \equiv \partial_x + y'\,\partial_y + \phi(x,y,y')\,\partial_{y'}\,,
\end{equation}
is a $S$-function associated with the 2ODE $\,y''=\phi(x,y,y')\,$.}

\bigskip
\noindent
To prove this result we will use the following lemma:

\bigskip
\bigskip
{\bf Lemma 1:} {\it Let $\,y''=\phi(x,y,y')\,$ be a 2ODE. If
$\,\overline{\eta}(x,y,y')\,$ is the infinitesimal of a Lie symmetry generator
in the evolutionary form, then $\,\overline{\eta}\,$ must obey the following PDE:
\begin{equation}
\label{spdesym}
D_x^2[\overline{\eta}]=D_x[\overline{\eta}]\,\phi_{y'}+\overline{\eta}\,\phi_y,
\end{equation}
\noindent
where $\,D_x\,$ is the operator defined by (\ref{defd}).}

\bigskip
\noindent
{\bf Proof of Lemma 1:} If the hypothesis of the lemma is fulfilled, then
\begin{equation}
\label{x2eq}
X^{(2)}[y'' - \phi(x,y,y')] = 0,
\end{equation}
where $\,X^{(2)}=\xi\,\partial_x+\eta\,\partial_y+\eta^{(1)}\,\partial_{y'}
+ \eta^{(2)}\,\partial_{y''}\,$
is the second extension of the group generator $\,X=\xi\,\partial_x+
\eta\,\partial_y\,$ and $\,\eta^{(1)}\,$ and $\,\eta^{(2)}\,$ are given by
\begin{eqnarray}
\label{eqeta1}
\eta^{(1)} &\equiv& D_x[\eta]-y'\,D_x[\xi], \\
\label{eqeta1}
\eta^{(2)} &\equiv& D_x[\eta^{(1)}]-y'\,D_x[\xi]=
D_x\left[D_x[\eta]-y'\,D_x[\xi]\right]-y'\,D_x[\xi].
\end{eqnarray}
\bigskip
\noindent
Since $\,X\,$ is a symmetry generator for the 2ODE
$\,y''=\phi(x,y,y')\,$ then the vector field $\,\hat{X}^{(1)}\,$ defined by
\begin{eqnarray}
\hat{X}^{(1)} &\equiv& X^{(1)} - \rho\,D_x\ = \nonumber \\ [.2cm]
&=& \xi\,\partial_x+\eta\,\partial_y+\eta^{(1)}\,\partial_{y'} -
\rho\,\partial_x-\rho\,y'\,\partial_y-\rho\,\phi\,\partial_{y'}= \nonumber \\ [.2cm]
&=& (\xi-\rho)\,\partial_x+(\eta-\rho\,y')\,\partial_y+
(\eta^{(1)}-\rho\,\phi)\,\partial_{y'}= \nonumber \\ [.2cm]
&=& (\xi-\rho)\,\partial_x+(\eta-y'\rho)\,\partial_y+
(D_x[\eta-y'\rho]-y'\,D_x[\xi-\rho])\,\partial_{y'},
\end{eqnarray}
\noindent
is also a symmetry generator. Choosing $\,\rho = \xi\,$ we obtain a
symmetry generator $\overline{X}^{(1)}$ in the evolutionary form:
\begin{equation}
\label{x1evol}
\overline{X}^{(1)}=\overline{\eta}\,\partial_y+
D_x[\overline{\eta}]\,\partial_{y'},
\end{equation}
\noindent
where $\,\overline{\eta} \equiv \eta-y'\xi\,$. So, we have that
$\overline{X}^{(2)}=\overline{\eta}\,\partial_y+
D_x[\overline{\eta}]\,\partial_{y'}+
D_x^2[\overline{\eta}]\,\partial_{y''}$ and we can write (\ref{x2eq}) as
\begin{equation}
\label{eqsym}
D_x^2[\overline{\eta}] - D_x[\overline{\eta}]\,\phi_{y'}-
\overline{\eta}\,\phi_y = 0,
\end{equation}
\bigskip
\noindent
as we want to demonstrate.$\Box$

\bigskip
Now we can prove theorem 1:

\bigskip
\noindent
{\bf Proof of Theorem 1:} If the hypothesis of the theorem is fulfilled, then
we have that (see section \ref{defs}, eq.(\ref{di4}))

$$
dI=R\left((\phi+\,S\,y')dx - S\,dy - dy'\right).
$$

\noindent
So, we can write
$
I_x = R\,(\phi+S\,y'), \,\,\, I_y = - S\,R,  \,\,\, I_{y'} = - R.
$
Using the compatibility conditions $(I_{xy}-I_{yx}=0, I_{xy'}-I_{y'x}=0$ and
$I_{yy'}-I_{y'y}=0 )$, we get:
\begin{eqnarray}
\label{condcomp1}
&&R_y\,(\phi+S\,y')+R\,(\phi_y+S_y\,y') + (S_x\,R + S\,R_x)=0\,,\\ [.3cm]
\label{condcomp2}
&&R_{y'}\,(\phi+S\,y')+R\,(\phi_{y'}+S_{y'}\,y'+S) + R_x=0\,, \\ [.3cm]
\label{condcomp3}
&&- ( R_{y'}\,S+R\,S_{y'}) + R_y=0\,.
\end{eqnarray}

\noindent
Eq.(\ref{condcomp2}) plus eq.(\ref{condcomp3}) times $y'$ results
\begin{equation}
\label{eqproof1}
R_x + y'\,R_y+\phi\,R_{y'}+R\,(\phi_{y'}+S) =0\,,
\end{equation}

\noindent
and eq.(\ref{condcomp1}) minus eq.(\ref{condcomp3}) times $\phi$ results
\begin{equation}
\label{eqproof2}
S\,(R_x + y'\,R_y+\phi\,R_{y'})+R\,(S_x + y'\,S_y+\phi\,S_{y'})+R\,\phi_y =0\,.
\end{equation}

\noindent
These equations can be written, respectively, as
\begin{eqnarray}
\label{eqproof1r}
&&D_x[R]+R\,(\phi_{y'}+S) =0\,,\\ [.3cm]
\label{eqproof2r}
&&S\,D_x[R]+R\,D_x[S]+R\,\phi_y =0\,.
\end{eqnarray}

\noindent
where $\,D_x\,$ is the operator defined in (\ref{defd}). Isolating $\,D_x\,[R]\,$
in eq.(\ref{eqproof1r}) and substituting in eq.(\ref{eqproof2r}) we have that
the $S$-function must obey the following equation:
\begin{equation}
\label{eqs}
D_x[S]=S^2+\phi_{y'}\,S-\phi_y\,.
\end{equation}

\noindent
Now, let $\,\overline{\eta}\,$ be an infinitesimal defining a Lie symmetry
in the evolutionary form. Substituting $S = - \frac{D_x[\overline{\eta}]}
{\overline{\eta}}$ in (\ref{eqs}) we have
\begin{equation}
\label{eqproof3}
D_x\left[-
\frac{D_x[\overline{\eta}]}{\overline{\eta}}\right]= \left(-
\frac{D_x[\overline{\eta}]}{\overline{\eta}}\right)^2+
\phi_{y'}\left(-
\frac{D_x[\overline{\eta}]}{\overline{\eta}}\right)- \phi_y\, \,\,
\Rightarrow\,\,\, -
\frac{D_x^2[\overline{\eta}]}{\overline{\eta}}= -
\,\phi_{y'}\frac{D_x[\overline{\eta}]}{\overline{\eta}}-\phi_y\,.
\end{equation}
\noindent
By using lemma 1 we can verify that eq.(\ref{eqproof3}) is an identity and the
theorem is demonstrated. $\Box$

\bigskip
If the reader is familiar with the classical theory of ODEs he/she can note two
interesting facts:

\begin{itemize}

\item Since the operator $\,D_x\,$ represents (over the solutions of the 2ODE)
the total derivative with respect to $x$ (i.e., $\frac{d}{dx}$), the PDE for
the $S$-function is formally a Riccati ODE\footnote{A Riccati ODE can be
generally written as $u'=f(x)\,u^2+g(x)\,u+h(x)$.}.

\item If we apply the transformation $u=-v'/v$ into the Riccati ODE
$u'=u^2+g(x)\,u-h(x)$ we get the linear 2ODE given by $v''=g(x)\,v'+h(x)\,v$.
This 2ODE, is formally analogous to the linear second order PDE (linear 2PDE)
for $\overline{\eta}$.

\end{itemize}

In the next section we will use the relation (\ref{sfunc}) to construct a
semi-algorithmic method to calculate the Lie symmetries of a rational 2ODE.
This method can be applied even to the case where there are only dynamical
symmetries.

\section{A method to calculate the Lie symmetries}
\label{calcls}

\hspace\parindent
The procedure can be divided in two main parts: first, we use the method
developed in \cite{Nosamc2007} to calculate the $S$-functions associated
with the 2ODE; then we use the relation (\ref{sfunc}) to calculate the
symmetries.

\subsection{Calculating the $S$-functions}
\label{calcsf}

\hspace\parindent
Based on theorem 2 of \cite{Nosamc2007} we have the following
result\footnote{For a proof, please see \cite{Nosamc2007}}:

\bigskip
{\bf Theorem 2: }{\it Let
\begin{equation}
\label{2ode2}
y'' = \phi(x,y,y') = \frac{M(x,y,y')}{N(x,y,y')},
\end{equation}
be a rational 2ODE (i.e., $M$ and $N$ are polynomial functions of $(x,y,y')$),
and Let $I(x,y,y')$ be an elementary first integral of it. Then, there is a
$S$-function $(S)$ associated with the 2ODE (\ref{2ode2}) such that:

\hspace\parindent
$(i)$ $S$ is an algebraic function of $(x,y,y')$.

\hspace\parindent
$(ii)$ The polynomial that defines the algebraic function $S$ is an
eigenpolynomial of the operator

${\cal D} \equiv (N)\,D + \left(N^2\,S^2+
(N\,\partial_{y'}M - M\,\partial_{y'}N)\,S+N\,\partial_{y}M - M\,\partial_{y}N\right)\partial_S$,

\noindent
where $\,{D} \equiv N\,D_x\,$ ($D_x\,$ is the operator defined by
(\ref{defd})).}

\bigskip
\noindent
The claiming $(ii)$ is simply a more formal way of saying that the $S$-function
must obey (\ref{eqs}). Besides, it turns clearer what to do to obtain the
$S$-function: we `only' have to calculate the eigenpolynomials (the Darboux
polynomials) of the operator
${\cal D}$. Each eigenpolynomial defines an algebraic function of $(x,y,y')$
which is a $S$-function associated with the 2ODE (\ref{2ode2}). The emphasis
on the word {\em only} is just to remember that calculating the eigenpolynomials
of the operator ${\cal D}$ may not be an easy task. But, although the procedure
may be hard to apply, it is of an algorithmic nature.

In this paper we want to deal with 2ODEs integrable by quadratures (i.e.,
presenting Liouvillian\footnote{For a formal definition of Liouvillian function,
please see \cite{Davenport}.} first integrals). Since we do not know the general
form of the $S$-funtions for this case, we will restrict ourselves to the
case where the 2ODE presents rational $S$-funtions. We will show that even
with this restriction, we can find symmetries in a lot of interesting cases.

\subsection{Finding the symmetries}
\label{findsym}

\hspace\parindent
Once we have found the rational $S$-functions associated with the 2ODE
(\ref{2ode2}), we will use the relation $D_x[\overline{\eta}]/\overline{\eta}
=-S$ to obtain a Lie symmetry. As we have mentioned in the last section, we
will restrict ourselves to the case where the 2ODE (\ref{2ode2}) presents
rational $S$-funtions. For this case we were able to prove the following
result (this paper's second goal):

\bigskip
{\bf Theorem 3:} {\it Let $\,y''=\phi(x,y,y')=M/N\,$ ($M$ and $N$ polynomials of
$\,(x,y,y')\,$) be a rational 2ODE presenting two independent Liouvillian first
integrals $\,I_1(x,y,y')\,$ and $\,I_2(x,y,y')\,$. If this 2ODE has two rational
$S$-functions $\,S_1\,$ and $\,S_2\,$ associated with it such that
$dI_1=R_1((\phi+\,S_1\,y')dx - S_1\,dy - dy')$ and
$dI_2=R_2((\phi+\,S_2\,y')dx - S_2\,dy - dy')$,
then the following statements hold:

\hspace\parindent
$(i)$ There exists
two independent Lie symmetries in the evolutionary form (\ref{x1evol}) such that
the infinitesimals $\,\overline{\eta}_1\,$ and $\,\overline{\eta}_2\,$ are
Darboux functions of $\,(x,y,y')\,$, i.e., the infinitesimals
$\,\overline{\eta}_1\,$ and $\,\overline{\eta}_2\,$ have the form
\begin{eqnarray}
\label{etafunc1}
\overline{\eta}_1 &=& e^{\frac{A_1}{B_1}}\,\prod_i {{p_1}_i}^{c_1}_i,
\\ [.3cm]
\label{etafunc2}
\overline{\eta}_2 &=& e^{\frac{A_2}{B_2}}\,\prod_j {{p_2}_j}^{c_2}_j.
\end{eqnarray}
\noindent
where $A_1$, $B_1$, $A_2$, $B_2$, the ${{p_1}_i}$ and the ${{p_2}_j}$ are
polynomial functions of $(x,y,y')$ and the ${c_1}_i$ and ${c_2}_j$ are
constants.

\hspace\parindent
$(ii)$ The ${{p_1}_i}$ and the ${{p_2}_j}$ are irreductible eigen-polynomials
(Darboux polynomials) of the operator $D$ (defined by $\,N\,D_x\,$) or are
irreductible factors of the denominators of $\,S_1,\,S_2\,$, respectively.

\hspace\parindent
$(iii)$ The irreductible factors of the polynomials $B_1$ and $B_2$ are
irreductible eigen-polynomials (Darboux polynomials) of the operator $D$ or are
irreductible factors of the denominators of $\,S_1,\,S_2\,$, respectively.}

\bigskip
\noindent
In order to prove this theorem we will need some results:

\bigskip
{\bf Proposition 1:} {\it Let $\,y''=\phi(x,y,y')\,$ be a 2ODE presenting a
first integral $\,I(x,y,y')\,$. If $\,S(x,y,y')\,$ is a $S$-function
associated with it, then the 1ODE defined by
\begin{equation}
\label{odeaux}
\frac{dv}{du} = - S(a_1,u,v)
\end{equation}
\noindent
has $\,I(a_1,u,v)=C\,$ (where $C$ is a constant) as its general solution.}

\bigskip
\noindent
{\bf Proof of Proposition 1:}
$\,I(a_1,u,v)=C\,$ is a general
solution of the 1ODE $dv/du= - S(a_1,u,v)$ if and only if
\begin{equation}
\label{dsig}
D_{u} [I] = 0,
\end{equation}
\noindent
where $D_{u} \equiv \partial_u - S\,\partial_v$. The condition
(\ref{dsig}) means that $\partial_u\,I - S\,\partial_v\,I=0$, i.e.,
$\,S=I_u/I_v\,$. Consider that the hypothesis of proposition 1 are
satisfied. To prove proposition 1 we only have to prove that
$\,S=I_y/I_{y'}\,$. According to (\ref{di4}) we can write
$dI=R((\phi+\,S\,y')dx - S\,dy - dy')$, leading to
$\,I_y= - R\,S\,$ and $\,I_{y'}= - R\,$. So, $\,S=I_y/I_{y'}\,$.
$\,\,\Box$

\bigskip
\noindent
Using similar reasoning we can enunciate the following\footnote{The proofs of
propositions 2 and 3 are analogous to that of proposition 1.}:

\bigskip
{\bf Proposition 2: }{\it Let $\,y''=\phi(x,y,y')\,$ be a 2ODE presenting a
first integral $\,I(x,y,y')\,$. If $\,S(x,y,y')\,$ is a $S$-function
associated with it, then the 1ODE defined by
\begin{equation}
\label{odeaux2}
\frac{dv}{dt} = \phi(t,a_2,v) + v\,S(t,a_2,v)
\end{equation}
\noindent
has $\,I(t,a_2,v)=C\,$ (where $C$ is a constant) as its general solution.}

\bigskip
{\bf Proposition 3: }{\it Let $\,y''=\phi(x,y,y')\,$ be a 2ODE presenting a
first integral $\,I(x,y,y')\,$. If $\,S(x,y,y')\,$ is a $S$-function
associated with it, then the 1ODE defined by
\begin{equation}
\label{odeaux3}
\frac{du}{dt} = \frac{\phi(t,u,a_3) + a_3\,S(t,u,a_3)}{S(t,u,a_3)}
\end{equation}
\noindent
has $\,I(t,u,a_3)=C\,$ (where $C$ is a constant) as its general solution.}

\bigskip
{\bf Definition 3:} {\it The three 1ODEs (\ref{odeaux}), (\ref{odeaux2}) and
(\ref{odeaux3}) will be called {\em auxiliary 1ODEs} of the 2ODE (\ref{2ode2})
associated with the $S$-function $\,S(x,y,y')\,$.}

\bigskip
In what follows, we will refer to (\ref{odeaux}) as 1ODE A1, to (\ref{odeaux2})
as 1ODE A2 and to (\ref{odeaux3}) as 1ODE A3.
This concept (of auxiliary 1ODE) will be useful to enunciate the following
result. Note that the general solution of the auxiliary 1ODEs (A1, A2 and A3)
are defined by the function $\,I\,$ that is a first integral of the 2ODE. The
following result establishes another important link between the 2ODE and the
auxiliary 1ODEs.

\bigskip
{\bf Proposition 4:} {\it Let $\,y''=\phi(x,y,y')\,$ be a 2ODE presenting a
Lie symmetry that can be written (in the evolutionary form) as
\begin{equation}
\label{x1evol2}
\overline{X}^{(1)}=\overline{\eta}(x,y,y')\,\partial_y+
D_x[\overline{\eta}(x,y,y')]\,\partial_{y'}.
\end{equation}
\noindent
Let $\,\overline{\zeta} \equiv D_x[\overline{\eta}]\,$, then the operator
$\,Y \equiv \overline{\eta}(a_1,u,v)\,\partial_u+
\overline{\zeta}(a_1,u,v)\,\partial_v$
is a generator for a Lie point symmetry of an auxiliary 1ODE A1
(\ref{odeaux}).}

\bigskip
\noindent
{\bf Proof of Proposition 4:}
The condition for a vector field
$\,Y \equiv \overline{\eta}(a_1,u,v)\,\partial_u+
\overline{\zeta}(a_1,u,v)\,\partial_v$ to be a symmetry vector of the 1ODE
(\ref{odeaux}) is that its commutator with the operator
$D_{u} \equiv \partial_u - S\,\partial_v$ is proportional to
$D_{u}$. So we have to prove that $[Y,D_{u}]$ is proportional to
$D_{u}$. We have that:
\begin{eqnarray}
\label{comydsig}
[Y,D_{u}] &=& -(\overline{\eta}\,S_u+
\overline{\zeta}\,S_v)\,\partial_v - (\overline{\eta}_u -
S\,\overline{\eta}_v)\,\partial_u - (\overline{\zeta}_u -
S\,\overline{\zeta}_v)\,\partial_v = \nonumber \\
&=& - D_{u}[\overline{\eta}]\,\partial_u
- (\overline{\eta}\,S_u + \overline{\zeta}\,S_v +
\overline{\zeta}_u - S\,\overline{\zeta}_v)\,\partial_v.
\end{eqnarray}
\noindent
Since $\,S=-D_x[\overline{\eta}]/\overline{\eta}=-
\overline{\zeta}/\overline{\eta}\,$ we can write
\begin{eqnarray}
\label{comydsig2}
[Y,D_{u}] &=& - D_{u}[\overline{\eta}]\,\partial_u
- \left( \overline{\eta}\,
\frac{\overline{\eta}\,\overline{\zeta}_u+\overline{\zeta}\,\overline{\eta}_u}
{\overline{\eta}^2}+
\overline{\zeta}\,
\frac{\overline{\eta}\,\overline{\zeta}_v+\overline{\zeta}\,\overline{\eta}_v}
{\overline{\eta}^2} +
\overline{\zeta}_u -
S\,\overline{\zeta}_v \right)\,\partial_v
\nonumber \\
&=& - D_{u}[\overline{\eta}]\,\partial_u
- (- S\,D_{u}[\overline{\eta}])\,\partial_v,
\end{eqnarray}
\noindent
implying that $\,[Y,D_{u}]= - D_{u}[\overline{\eta}]\,D_{u}$.
$\,\,\Box$

\bigskip
{\bf Proposition 5: }{Let $\,y''=\phi(x,y,y')\,$ be a 2ODE presenting two
independent first integrals $\,I_1(x,y,y')\,$ and $\,I_2(x,y,y')\,$. If
$\,S_1 = {I_1}_y/{I_1}_z\,$ and $\,S_2 = {I_2}_y/{I_2}_z\,$ are
two $S$-functions associated with it such that
$dI_1=R_1((\phi+\,S_1\,y')dx - S_1\,dy - dy')$ and
$dI_2=R_2((\phi+\,S_2\,y')dx - S_2\,dy - dy')$, then there are
infinitesimals for Lie symmetries in the evolutionary form,
$\,\overline{\eta}_1\,$ and $\,\overline{\eta}_2\,$, (such that
$\,-D_x[\overline{\eta}_1]/\overline{\eta}_1=S_1\,$ and
$\,-D_x[\overline{\eta}_2]/\overline{\eta}_2=S_2\,$) given by
\begin{eqnarray}
\label{eta1format}
\overline{\eta}_1 = \frac{1}{(S_1-S_2)\,R_2}, \\
\label{eta2format}
\overline{\eta}_2 = \frac{1}{(S_2-S_1)\,R_1}.
\end{eqnarray}
}

\bigskip
\noindent
{\bf Proof of Proposition 5: }
Consider that the hypothesis of the proposition are fulfilled. Let the
operators $X_1 \equiv \overline{\eta}_1\,\partial_y+
D_x[\overline{\eta}_1]\,\partial_{y'}$ and
$X_2 \equiv \overline{\eta}_2\,\partial_y+D_x[\overline{\eta}_2]\,\partial_{y'}$
be Lie symmetries of the 2ODE $\,y''=\phi(x,y,y')\,$ in the evolutionary form.
Without loss of generality we can choose $\overline{\eta}_1$ and
$\overline{\eta}_2$ such that $\,X_1[I_1]=X_2[I_2]=0\,$ and
$\,X_1[I_2]=X_2[I_1]=1\,$. Using $\,X_1[I_1]=0\,$ and $\,X_1[I_2]=1\,$ we have
\begin{eqnarray}
\label{eta1formproof}
\overline{\eta}_1\,{I_1}_y+D_x[\overline{\eta}_1]\,{I_1}_{y'}=0, \nonumber \\
\label{eta2formproof}
\overline{\eta}_1\,{I_1}_y+D_x[\overline{\eta}_1]\,{I_2}_{y'}=1, \nonumber
\end{eqnarray}
\noindent
implying that
\begin{equation}
\label{eta1proof}
\overline{\eta}_1 = \frac{- {I_1}_{y'}}
{{I_1}_{y}\,{I_2}_{y'}-{I_1}_{y'}\,{I_2}_{y}} =
\frac{- 1}
{\frac{{I_1}_{y}}{{I_1}_{y'}}\,{I_2}_{y'}-{I_2}_{y}} =
\frac{- 1}
{{I_2}_{y'}\left(\frac{{I_1}_{y}}{{I_1}_{y'}}-
\frac{{I_2}_{y}}{{I_2}_{y'}}\right)}\,.
\end{equation}
\noindent
Since $\,{I_2}_{y'}=-R_2\,$, $\,{I_1}_{y}/{I_1}_{y'}=S_1\,$ and
$\,{I_2}_{y}/{I_2}_{y'}=S_2\,$ we can write $\,\overline{\eta}_1=
1/((S_1-S_2)\,R_2)\,$ and we can deduce the format (\ref{eta2format})
for $\,\overline{\eta}_2\,$ in an analogous way.$\,\,\Box$

\bigskip
{\bf Theorem 4:} {\it If a rational 1ODE written in the form
$\,y'=M(x,y)/N(x,y)\,$ ($\,M\,$ and $\,N\,$ are polynomials with no common
factor) has a general solution of the form $\,I(x,y)=C\,$ where $\,I\,$ is a
Liouvillian function of its arguments and $\,C\,$ is an arbitrary constant,
then the 1-form $\,M\,dx-N\,dy\,$ presents an integrating factor $\,R\,$
of the form
\begin{equation}
\label{Rformat}
R = e^{r_0}\prod_i\,{p_i}^{c_i},
\end{equation}
\noindent
where $r_0$ is a rational function of $(x,y)$, the $p_i$ and the factors of the
denominator of $r_0$ are irreductible Darboux polynomials of the operator
$\,N\,\partial_x+M\,\partial_y\,$ and the $c_i$ are constants.
}

\bigskip
\noindent
For a proof of theorem 4 see
\cite{Nosjpa2002,Noscpc2002,Nosjpa2002-2,Nosjcam2004} or \cite{Christopher,Christopher2}.

\bigskip

Now we can prove theorem 3.  The key point in the following demonstration is
the fact that the general solutions of the auxiliary 1ODEs associated with
$\,S\,$ are defined by the same function $\,I\,$ that is a first integral of
the rational 2ODE $\,y''=\phi(x,y,y')=M/N\,$.

\bigskip

\noindent
{\bf Proof of Theorem 3: }
First of all let's stablish some notation: we will write $\,S_1 = P_1/Q_1\,$
and $\,S_2 = P_2/Q_2\,$, where $\,P_1\,$, $\,Q_1\,$, $\,P_2\,$ and $\,Q_2\,$
are polynomials of $\,(x,y,y')\,$ and the pairs $\,P_1\,$, $\,Q_1\,$ and
$\,P_2\,$, $\,Q_2\,$ do not have any common factors.

\bigskip
\noindent
Consider now that the hypothesis of the theorem are fulfilled.
From proposition 1, we have that $\,I_1(a_1,u,v)=C\,$ is a general Liouvillian
solution of the rational auxiliary 1ODE A1 associated with $\,S_1\,$
\begin{equation}
\label{1odea1s1}
\frac{dv}{du}= - S_1(a_1,u,v) = -\frac{P_1(a_1,u,v)}{Q_1(a_1,u,v)}.
\end{equation}
\noindent
Then (from theorem 4) the 1-form $\,P_1(a_1,u,v)\,du+Q_1(a_1,u,v)\,dv\,$
presents an integrating factor of the form (\ref{Rformat}). Let $\,R_{[A1,S1]}(a_1,u,v)\,$ be such an integrating factor. Since
\begin{eqnarray}
dI_1(x,y,y')&=&R_1\left((\phi+\,S_1\,y')dx - S_1\,dy - dy'\right)
\nonumber \\ [.3cm]
\label{dI1p3}
&=& \frac{R_1}{Q_1\,N}\,\left((M\,Q_1+\,P_1\,N\,y')dx -
P_1\,N\,dy - Q_1\,N\,dy'\right),
\end{eqnarray}
we can see that
$$
\frac{\partial I_1}{\partial y}(x,y,y')=-\frac{R_1(x,y,y')\,P_1(x,y,y')}
{Q_1(x,y,y')} \,\,\,\,{\rm and}\,\,\,\, \frac{\partial I_1}{\partial {y'}}
(x,y,y')=-R_1(x,y,y')
$$
implying that
$$
\frac{\partial I_1}{\partial u}(a_1,u,v)=
-\frac{R_1(a_1,u,v)\,P_1(a_1,u,v)}{Q_1(a_1,u,v)} \,\,\,{\rm and}\,\,\,
\frac{\partial I_1}{\partial v}(a_1,u,v) = -R_1(a_1,u,v).
$$
Therefore, $\,- R_1(a_1,u,v)/Q_1(a_1,u,v)\,$ is also an integrating factor
for the 1-form $\,P_1(a_1,u,v)\,du+Q_1(a_1,u,v)\,dv\,$. Then, we can write
$\,\frac{R_1}{Q_1}={\cal F}_1(I_1)\,R_{[A1,S1]}\,$
(where $\,{\cal F}_1\,$ is a function of $\,I_1\,$) or
$\,\frac{R_1}{Q_1}=k_1\,R_{[A1,S1]}\,$
(where $\,k_1\,$ is a constant) leading to
$\,R_1={\cal F}_1(I_1)\,R_{[A1,S1]}\,Q_1\,$
(or $\,R_1=k_1\,R_{[A1,S1]}\,Q_1\,$). So,
$\,R_{[A1,S1]}(x,y,y')\,Q_1(x,y,y')\,$ is an integrating factor for the
1-form $\,(\phi+\,S_1\,y')dx - S_1\,dy - dy'\,$.

\bigskip
\noindent
Besides, from proposition 2 we have that $\,I_1(t,a_2,v)=C\,$ is a general
Liouvillian solution of the rational auxiliary 1ODE A2 associated with $\,S_1\,$
\begin{equation}
\label{1odea2s1}
\frac{dv}{dt} = \phi(t,a_2,v) + v\,S_1(t,a_2,v)
= \frac{Q_1(t,a_2,v)\,M(t,a_2,v)+ v\,P_1(t,a_2,v)\,N(t,a_2,v)}
{Q_1(t,a_2,v)\,N(t,a_2,v)}.
\end{equation}
\noindent
Then (from theorem 4) the 1-form $\,-(Q_1\,M+ v\,P_1\,N)\,dt+Q_1\,N\,dv\,$
presents an integrating factor of the form (\ref{Rformat}). Let
$\,R_{[A2,S1]}(t,a_2,v)\,$ be such an integrating factor. From (\ref{dI1p3})
and using a completely analogous reasoning of what we have shown above, we can
conclude that $\,R_1(t,a_2,v)/\left(Q_1(t,a_2,v)\,N(t,a_2,v)\right)\,$ is also
an integrating factor for the 1-form
$\,-(Q_1\,M+ v\,P_1\,N)\,dt+Q_1\,N\,dv\,$. Then, we can write
$\,\frac{R_1}{Q_1\,N}={\cal F}_2(I_1)\,R_{[A2,S1]}\,$
(where $\,{\cal F}_2\,$ is a function of $\,I_1\,$) or
$\,\frac{R_1}{Q_1\,N}=k_2\,R_{[A2,S1]}\,$
(where $\,k_2\,$ is a constant) leading to
$\,R_1={\cal F}_2(I_1)\,R_{[A2,S1]}\,Q_1\,N\,$
(or $\,R_1=k_2\,R_{[A2,S1]}\,Q_1\,N\,$). So,
$\,R_{[A2,S1]}(x,y,y')\,Q_1(x,y,y')\,N(x,y,y')\,$ is an integrating factor for
the 1-form $\,(\phi+\,S_1\,y')dx - S_1\,dy - dy'\,$.

\bigskip
Now let's see: since $\,R_{[A1,S1]}(x,y,y')\,Q_1(x,y,y')\,$
and $\,R_{[A2,S1]}(x,y,y')\,Q_1(x,y,y')$
$N(x,y,y')\,$
are integrating factors for the 1-form
$\,(\phi+\,S_1\,y')dx - S_1\,dy - dy'\,$, then we can write  $\,R_{[A1,S1]}(x,y,y')Q_1(x,y,y')=
{\cal F}(I_1)\,R_{[A2,S1]}(x,y,y')\,Q_1(x,y,y')N(x,y,y')\,$
($\,{\cal F}\,$ is a function of $\,I_1\,$) or
$\,R_{[A1,S1]}(x,y,y')\,Q_1(x,y,y')=
k_1\,R_{[A2,S1]}(x,y,y')\,Q_1(x,y,$
$y')\,N(x,y,y')\,$ ($\,k\,$ is a constant).

\begin{itemize}
\item First possibility:
$\,R_{[A1,S1]}(x,y,y')=k\,R_{[A2,S1]}(x,y,y')\,N(x,y,y')\,$.
Let's remember that the polynomials forming
$\,R_{[A1,S1]}(x,y,y')\,$ are polynomials of $\,(y,y')\,$ and the polynomials
forming $\,R_{[A2,S1]}(x,y,y')\,$ are polynomials of
$\,(x,y')\,$. Therefore, we can conclude that $\,R_{[A1,S1]}(x,y,y')\,$ and
$\,R_{[A2,S1]}(x,y,y')\,$ are of the form $\,\exp[r_0]\,\prod_i\,{p_i}^{c_i}\,$,
where the $\,p_i\,$ are polynomials of $\,(x,y,y')\,$ and $\,r_0\,$ is a
rational function of $\,(x,y,y')\,$.
\item Second possibility:
$\,R_{[A1,S1]}(x,y,y')={\cal F}(I_1)\,R_{[A2,S1]}(x,y,y')\,N(x,y,y')\,$.
For the case where $\,I_1\,$ is an elementary function of $\,(x,y,y')\,$, from
the results presented in \cite{Nosjmp2009} we can conclude that
$\,R_{[A1,S1]}(x,y,y')\,$ and $\,R_{[A2,S1]}(x,y,y')\,$ are of the form
$\,\prod_i\,{p_i}^{c_i}\,$ and use the same reasoning above to show that the
$\,p_i\,$ are polynomials of $\,(x,y,y')\,$.

Now, let's suppose that $\,I_1\,$ is a non elementary Liouvillian function of
$\,(x,y,y')\,$. Since $\,R_{[A1,S1]}(x,y,y')\,$ and $\,R_{[A2,S1]}(x,y,y')\,$
are elementary functions, the only possibility for the relation
$\,R_{[A1,S1]}(x,y,y')={\cal F}(I_1)\,R_{[A2,S1]}(x,y,y')\,N(x,y,y')\,$ to be
true is that $\,{\cal F}(I_1)\,$ is the constant function, i.e.,
$\,{\cal F}(I_1)=k\,$. So, we are again in the first possibility and we have
that $\,R_{[A1,S1]}(x,y,y')\,$ and $\,R_{[A2,S1]}(x,y,y')\,$ are of the form
$\,\exp[r_0]\,\prod_i\,{p_i}^{c_i}\,$, where the $\,p_i\,$ are polynomials of
$\,(x,y,y')\,$ and $\,r_0\,$ is a rational function of $\,(x,y,y')\,$.
\end{itemize}

Now, consider the auxiliary rational 1ODEs A1 and A2 associated with $\,S_2\,$:
\begin{equation}
\label{1odea1s2}
\frac{dv}{du} = - S_2(a_1,u,v) = -\frac{P_2(a_1,u,v)}{Q_2(a_1,u,v)},
\end{equation}
\begin{equation}
\label{1odea2s2}
\frac{dv}{dt} = \phi(t,a_2,v) + v\,S_2(t,a_2,v)
= \frac{Q_2(t,a_2,v)\,M(t,a_2,v)+ v\,P_2(t,a_2,v)\,N(t,a_2,v)}
{Q_2(t,a_2,v)\,N(t,a_2,v)}.
\end{equation}
\noindent
From theorem 4 (and using propositions 2 and 3 again), we can infer that the
1-forms $\,P_2(a_1,u,v)\,du+Q_2(a_1,u,v)\,dv\,$ and
$\,-(Q_2(t,a_2,v)\,M(t,a_2,v)+ v\,P_2(t,a_2,v)\,N(t,a_2,v$
$))\,dt+Q_2(t,a_2,v)\,N(t,a_2,v)\,dv\,$ present integrating factors of the form
(\ref{Rformat}). Let $\,R_{[A1,S2]}(a_1,u,v)\,$ and $\,R_{[A2,S2]}(t,a_2,v)\,$
be, respectively, these integrating factors. Using the same reasoning above we
can conclude that (analogously as above):
\begin{itemize}
\item $\,-R_{[A1,S2]}(x,y,y')\,Q_2(x,y,y')\,$ is an integrating factor for the
1-form $\,(\phi+\,S_2\,y')dx - S_2\,dy - dy'\,$.
\item $\,R_{[A2,S2]}(x,y,y')\,Q_2(x,y,y')\,N(x,y,y')\,$ is an integrating factor
for the 1-form $\,(\phi+\,S_2\,y')dx - S_2\,dy - dy'\,$.
\item $\,R_{[A1,S2]}(x,y,y')\,$ and $\,R_{[A2,S2]}(x,y,y')\,$ are of the form
$\,\exp[r_0]\,\prod_i\,{p_i}^{c_i}\,$, where the $\,p_i\,$ are polynomials of
$\,(x,y,y')\,$ and $\,r_0\,$ is a rational function of $\,(x,y,y')\,$.
\end{itemize}

Therefore, from these results and the analogous above, we can afirm that there
exists two independent Liouvillian first integrals $\,\overline{I}_1\,$ and
$\,\overline{I}_2\,$ such that
\begin{eqnarray}
\label{dI1over}
d\overline{I}_1(x,y,y')&=&\overline{R}_1\left((\phi+\,S_1\,y')dx -
S_1\,dy - dy'\right),
 \\ [.3cm]
\label{dI2over}
d\overline{I}_2(x,y,y')&=&\overline{R}_2\left((\phi+\,S_2\,y')dx -
S_2\,dy - dy'\right),
\end{eqnarray}
where $\,\overline{R}_1 \equiv -R_{[A1,S1]}\,Q_1\,$ and
$\,\overline{R}_2 \equiv -R_{[A1,S2]}\,Q_2\,$ are of the form
$\,e^{A/B}\,\prod_i\,{p_i}^{c_i}\,$ ($A,\,B,\,p_i\,$ polynomials
in $\,(x,y,y')\,$ and $\,{c_i}\,$ constants).

We can now prove $(i)$. From (\ref{dI1over}), (\ref{dI2over}) and proposition 5
we can conclude (since $\,S_1\,$ and $\,S_2\,$ are rational functions of
$\,(x,y,y')\,$) that there are infinitesimals $\,\overline{\eta}_1\,$ and
$\,\overline{\eta}_2\,$ of the form (\ref{etafunc1}) and (\ref{etafunc2}).

In order to prove $(ii)$ and $(iii)$ let's first note that
\begin{eqnarray}
\label{Deta1}
- \frac{D_x[\overline{\eta}_1]}{\overline{\eta}_1}=\frac{P_1}{Q_1}
\,\,\,\Rightarrow\,\,\,Q_1\frac{D[\overline{\eta}_1]}{\overline{\eta}_1}=
- N\,P_1\,, \\
\label{Deta2}
- \frac{D_x[\overline{\eta}_2]}{\overline{\eta}_2}=\frac{P_2}{Q_2}
\,\,\,\Rightarrow\,\,\,Q_2\frac{D[\overline{\eta}_2]}{\overline{\eta}_2}=
- N\,P_2\,,
\end{eqnarray}
\noindent
i.e., $\,Q_1\frac{D[\overline{\eta}_1]}{\overline{\eta}_1}\,$ and
$\,Q_2\frac{D[\overline{\eta}_2]}{\overline{\eta}_2}\,$ are polynomials in
$\,(x,y,y')\,$. So, we have the situation: the infinitesimals
$\,\overline{\eta}_1\,$ and $\,\overline{\eta}_2\,$ are Darboux functions such
that $\,\frac{{\cal D}_1[\overline{\eta}_1]}{\overline{\eta}_1}=pol_1\,$ and
$\,\frac{{\cal D}_2[\overline{\eta}_2]}{\overline{\eta}_2}=pol_2\,$, where
$\,{\cal D}_1 \equiv Q_1\,D\,$ and $\,{\cal D}_2 \equiv Q_2\,D\,$. Finally,
from the main result presented in \cite{Nosjpa2002-2}, we can directly conclude
$\,(ii)\,$ and $\,(iii)\,$.$\,\,\Box$

\bigskip
\bigskip
The procedure to calculate the symmetries in this case is
completely analogous to that for calculate the integrating factors
of rational 1ODEs with Liouvillian solution (see
\cite{Nosjcam2004}): briefly, we calculate the Darboux polynomials
(up to a certain degree) of the operator $\,D\,$. Then we
substitute the candidates for $\,\overline{\eta}\,$ in the
equation
\begin{equation}
\label{eqdeta}
Q\frac{D[\overline{\eta}]}{\overline{\eta}}=
- N\,P\,,
\end{equation}
\noindent
and we solve the resulting linear equations for the constants $\,c_i\,$.

\section{Examples}
\label{examples}

\hspace\parindent
In this section we will present two examples where we can use the
semi-algorithm presented above to calculate the Lie symmetries. In the first
example we will show a 2ODE with one point symmetry and one Darboux dynamical
symmetry to ilustrate theorem 3. This symmetry can be found by our method --
a process devoided of any guess. We will use this example to show our procedure
in action, i.e., we will provide comments about the calculations to clearer the
steps of the method. In the second example we use a 2ODE not presenting Lie
point symmetries but with two rational dynamical symmetries. The good point to
be noted is that, for our kind of process, the existence or not of point
symmetries makes no difference.

\subsection{First example}
\label{fexemp}

Consider the 2ODE given by
\begin{equation}
\label{eqex1}
y'' = \frac {2\,y-3\,zy+{z}^{2}y-zx+{z}^{2}x}{y ( y-x ) }.
\end{equation}
\noindent
To use the procedure explained in the section \ref{calcls} we have, first, to
obtain the operators $\,D\,$ and $\,{\cal D}\,$. They are given by
\begin{equation}
\label{Dex1}
D = (x\,y-{y}^{2}){\frac {\partial }{\partial x}} + (x\,y\,z-{y}^{2}z)
{\frac {\partial }{\partial y}} + ( 3\,y\,z-y{z}^{2}-x{z}^{2}-2\,y+x\,z)
{\frac {\partial }{\partial z}},
\end{equation}
$$
{\cal D} = ({y}^{4}+{y}^{2}{x}^{2}-2\,{y}^{3}x)
{\frac {\partial }{\partial x}} +
( -2\,z{y}^{3}x+z{y}^{4}+z{y}^{2}{x}^{2})
{\frac {\partial }{\partial y}} +
$$
$$
(yz{x}^{2}-y{x}^{2}{z}^{2}+2\,{y}^{3}+2\,{y}^{2}zx+{y}^{3}{z}^{2}-
3\,z{y}^{3}-2\,x{y}^{2}) {\frac {\partial }{\partial z}} +
$$
$$
\left( ( {y}^{2}{x}^{2}+{y}^{4}-2\,{y}^{3}x )\, {s}^{2}+ ( -2\,zy{x}^{2}+2\,z{y}^{3}+2\,x{y}^{2}+y{x}^{2}-3\,{y}^{3})\, s \right.
$$
\begin{equation}
\label{Dsex1}
\left. -{z}^{2}{x}^{2}+z{x}^{2}+{z}^{2}{y}^{2}-2\,zyx+2\,{y}^{2}-3\,z{y}^{2}+2\,{z
}^{2}yx
\right)
{\frac {\partial }{\partial s}}.
\end{equation}
\noindent
The second step is to calculate the Darboux polynomials of the operator
$\,{\cal D}\,$. We can find two Darboux polynomials:
\begin{equation}
\label{Darbs1_1}
{p_s}_1 = ( z-1 ) ( y+x ) + s\,y\,( x-y ) ,
\end{equation}
\begin{equation}
\label{Darbs2_1}
{p_s}_2 = x \,(-2\,y+zy+zx)(z-1)+s\,y\,(x-y)(zx-y).
\end{equation}

\noindent
The $\,S$-functions are the solutions of {\it Darboux polynomial of
$\,{\cal D}\,$= 0}. From (\ref{Darbs1_1}) and (\ref{Darbs2_1}) we can obtain
two rational $\,S$-functions given by
\begin{equation}
\label{S1ex1}
S1 = {\frac { \left( z-1 \right)  \left( y+x \right) }
{y \left( x-y \right) }},
\end{equation}
\begin{equation}
\label{S2ex1}
S2 = {\frac {x \left( -2\,y+zy+zx \right)  \left( z-1 \right) }
{ \left( x-y \right)  \left( zx-y \right) y}}.
\end{equation}

\noindent
The next step is to calculate the Darboux polynomials of the $\,{D}\,$ operator
and the corresponding cofactors to build the infinitesimals. They are
\begin{equation}
\label{Darb1_1}
p_1=y \,\,\,\,\Rightarrow \,\,\,\, q_1=\left( x-y \right) z,
\end{equation}
\begin{equation}
\label{Darb2_1}
p_2=x-y \,\,\,\,\Rightarrow \,\,\,\,
q_2=- \left( z-1 \right)\, y,
\end{equation}
\begin{equation}
\label{Darb3_1}
p_3=z-1 \,\,\,\,\Rightarrow \,\,\,\,
q_3=2\,y-zy-zx.
\end{equation}

\noindent
Using these Darboux polynomials and looking at (\ref{eqdeta}), we can use the
semi-algorithm described in \cite{Nosjcam2004,Noscpc2007} to calculate the
infinitesimals:
\begin{equation}
\label{eta1ex1}
\overline{\eta}_1=- \frac{\left( x-y \right) ^{2}{e^{{\frac {x-y}
{ \left( z-1 \right) y}}}}}{y},
\end{equation}
\begin{equation}
\label{eta2ex1}
\overline{\eta}_2={\frac { \left( zx-y \right)  \left( x-y \right) ^{2}}
{ \left( z-1 \right) y}}.
\end{equation}

\bigskip
\bigskip
We can use the relations $\,X_1[I_1]=0\,$, $\,X_2[I_1]=1\,$, $\,D[I_1]=0\,$ and
solve them to $\,{I_1}_x\,$, $\,{I_1}_y\,$ and $\,{I_1}_z\,$. We can also use
the relations $\,X_1[I_2]=1\,$, $\,X_2[I_2]=0\,$, $\,D[I_2]=0\,$ and
solve them to $\,{I_2}_x\,$, $\,{I_2}_y\,$ and $\,{I_2}_z\,$. From the
derivatives we can integrate and obtain $\,I_1\,$ and $\,I_2\,$:
\begin{equation}
\label{I1ex1}
I_1={\frac {y \left( z-1 \right) }{ \left( y-x \right) ^{2}}},
\end{equation}
\begin{equation}
\label{I2ex1}
I_2=\int \!\frac{{e^{{\frac {y-x}{y \left( z-1 \right) }}}}
\left( -2\,y+zy+zx \right) \,y}{\left( y-x \right)^3}\,dx.
\end{equation}

\bigskip

\subsection{second example}
\label{sexemp}

It can be proved (see \cite{MuRo}) that the 2ODE
\begin{equation}
\label{eqex2}
y'' = - \frac{x^2+4\,y^4+2\,y^2}{4\,y^3}
\end{equation}
\noindent
has no Lie point symmetries. By doing the procedure explained in the section
\ref{calcls} we have:

\noindent
The operator $\,D\,$ and $\,{\cal D}\,$ are given by
\begin{equation}
\label{Dex2}
D = 4\,{y}^{3}{\frac {\partial }{\partial x}} +4\,{y
}^{3}z{\frac {\partial }{\partial y}} + \left( -
{x}^{2}-4\,{y}^{4}-2\,{y}^{2} \right) {\frac {\partial }{\partial z}},
\end{equation}
$$
{\cal D} = 16\,{y}^{6}{\frac {\partial }{\partial x}} +16
\,{y}^{6}z{\frac {\partial }{\partial y}} +4\,
{y}^{2} \left( -4\,{y}^{5}-2\,{y}^{3}-y{x}^{2} \right) {\frac {
\partial }{\partial z}} +
$$
\begin{equation}
\label{Dsex2}
+4\,{y}^{2} \left( -3
\,{x}^{2}+4\,{y}^{4}{s}^{2}-2\,{y}^{2}+4\,{y}^{4} \right) {\frac {
\partial }{\partial s}}.
\end{equation}
\noindent
We can find two Darboux polynomials of the operator $\,{\cal D}\,$:
\begin{equation}
\label{Darbs1_2}
{p_s}_1 = x+yz + S\,{y}^{2},
\end{equation}
$$
{p_s}_2 = {4\,{y}^{4}x+8\,{y}^{5}z+4\,{y}^{2}x+4\,{y}^{3
}z+{x}^{3}+2\,{x}^{2}yz+2\,i\,({x}^{2}{y}^{2}+4\,{y}^{6})} +
$$
\begin{equation}
\label{Darbs2_2}
+ 4\,S\,{y}^{3}\,( -y+zx+2\,y{z}^{2}+2\,i\,{y}^{2}z ).
\end{equation}

\noindent
From (\ref{Darbs1_2}) and (\ref{Darbs2_2}) we can obtain two rational
$\,S$-functions given by
\begin{equation}
\label{S1ex2}
S1 = -{\frac {x+yz}{{y}^{2}}},
\end{equation}
\begin{equation}
\label{S2ex2}
S2 = {\frac {4\,{y}^{4}x+8\,{y}^{5}z+4\,{y}^{2}x+4\,{y}^{3
}z+{x}^{3}+2\,{x}^{2}yz+2\,i\,({x}^{2}{y}^{2}+4\,{y}^{6})}
{ 4\,\left( -y+zx+2\,y{z}^{2}+2\,i\,{y}^{2}z \right) {y}^{3}}}.
\end{equation}

\noindent
The Darboux polynomials of the $\,{D}\,$ operator and the corresponding
cofactors are
\begin{equation}
\label{Darb1_2}
p_1=y \,\,\,\,\Rightarrow \,\,\,\, q_1=4\,{y}^{2}z,
\end{equation}
\begin{equation}
\label{Darb2_2}
p_2=x+2\,yz+2\,i\,y^2 \,\,\,\,\Rightarrow \,\,\,\,
q_2=-2\,y( x-2\,yz-2\,i\,{y}^{2}),
\end{equation}
\begin{equation}
\label{Darb3_2}
p_3=x+2\,yz-2\,i\,y^2 \,\,\,\,\Rightarrow \,\,\,\,
q_3=-2\,y( x-2\,yz+2\,i\,{y}^{2}).
\end{equation}

\noindent
Using these Darboux polynomials, we can use (\ref{eqdeta}) to calculate the
infinitesimals
\begin{equation}
\label{eta1ex2}
\overline{\eta}_1={\frac {4\,{y}^{3}}{ \left( x+2\,yz-2\,i\,y^2 \right)
\left( x+2\,yz+2\,i\,y^2 \right) }},
\end{equation}
\begin{equation}
\label{eta2ex2}
\overline{\eta}_2=-\,{\frac {-y+zx+2\,y{z}^{2}+2\,i\,{y}^{2}z}
{4(x+2\,yz+2\,i\,y^2)}}.
\end{equation}

\bigskip
We can use the relations $\,X_1[I_1]=0\,$, $\,X_2[I_1]=1\,$, $\,D[I_1]=0\,$,
$\,X_1[I_2]=1\,$, $\,X_2[I_2]=0\,$, $\,D[I_2]=0\,$ to obtain $\,I_1\,$ and
$\,I_2\,$:
\begin{equation}
\label{I1ex2}
I_1=4\,x-2\,i\,\ln\left({\frac{x+2\,yz-2\,i\,{y}^{2}}
{x+2\,yz+2\,i\,{y}^{2}}}\right),
\end{equation}
\begin{equation}
\label{I2ex2}
I_2=\frac{1}{2}\,{y}^{2}+\frac{1}{2}\,i\,x+\ln(y)-\frac{1}{2}\,
\ln(x+2\,yz+2\,i\,{y}^{2}) -\frac{1}{8}\,{\frac{{x}^{2}}{{y}^{2}}}+
\frac{1}{2}\,{z}^{2}.
\end{equation}

\bigskip
In \cite{MuRo} Muriel and Romero were capable of dealing with this 1ODE
(\ref{eqex2}) by developing the concept of $\,C^{\,\infty}(M^{(1)})$-symmetries
-- vector fields (also called $\,\lambda$-symmetries) that are neither Lie
symmetries nor Lie-B\"{a}cklund symmetries. The determining equations for the
components of these vector fields depend on an arbitrary function $\,\lambda\,$,
which can be chosen in order to simplify the process of solution of the
determining equations. For this 1ODE they obtained $\,\lambda=x/u^2\,$ and
$\,v=x\frac{\partial}{\partial x}\,$ for the $\,\lambda$-symmetry.

In our process (shown above) the main cost (computational cost) is to calculate
the Darboux polynomials of the operator $\,{\cal D}\,$ (followed by calculating
the Darboux polynomials of the operator $\,D$). The good news are that the
process is entirely computational, i.e., devoid of any guess.

\section{Conclusion}
\label{conclu}

Although the Lie symmetry method is the most powerful method for solving,
reducing and studying dynamical systems since its appearance (end of XIX$^{th}$
century), there are (still today) a bunch of open questions. The main `gap' is
the absence of an algorithm to calculate the symmetry vector fields. The things
go worse when the dynamical system under study does not present Lie point
symmetries (in which case we can not even count with a systematic method to
deal with the problem).

In this paper we have proposed a semi-algorithm to find Lie symmetries for a
rational 2ODE. We began by showing a deep connection between the $S$-functions
(see section \ref{defs} and \cite{Nosjpa2001,Nosamc2007}) of a 2ODE and its
symmetries in the evolutionary form. Then, restricting ourselves to the case
where the $S$-functions are rational and the first integrals are Liouvillian,
we could develop a semi-algorithm to calculate the symmetries.

The great advantage of our method is that it converts the search for symmetries
into (essentially) searching the Darboux polynomials of the polynomial
differential linear operators $\,{\cal D}\,$ and $\,{D}\,$ (in other words,
into solving second degree algebraic equations) -- a semi-algorithmic procedure.
The disadvantage is that, as the degree of the Darboux polynomials grows, the
computational coast increases `a lot' and, sometimes, turns the entire process
inviable.

Our method is not general since it is limited to the cases where
the $S$-functions are rational (see section \ref{calcls}). However,
from a practical point of view, this restriction does not seems to be a great
one: all the non-linear rational 2ODEs (that we have analysed)
with two Liouvillian first integrals, presented rational $S$-functions.

In regard to future work, there are some main directions (open questions):
\begin{itemize}
\item What is the general form of the $S$-functions for a 2ODE presenting two
Liouvillian first integrals? (and only one Liouvillian first integral?)
\item What about NODEs?
\item What happens if we consider elementary functions present in the ODE?
\item What about systems of ODEs?
\item Etc...
\end{itemize}


\end{document}